\documentclass[11pt,a4paper]{article}
\usepackage[margin=2.6cm]{geometry}
\usepackage{amsmath,amssymb,bm,booktabs,array}
\usepackage{slashed}
\usepackage{tikz}
\usetikzlibrary{decorations.pathmorphing,decorations.markings,arrows.meta}
\usepackage[colorlinks=true,linkcolor=blue!60!black,citecolor=blue!60!black,urlcolor=blue!60!black]{hyperref}

\newcommand{\Tr}{\operatorname{Tr}}
\newcommand{\tr}{\operatorname{tr}}

\title{\textbf{Anomalies and the Berry Phase}\\[0.4em]
\large From $\pi^0\to\gamma\gamma$ to the geometry of the fermion vacuum\\[0.6em]
\normalsize\emph{Based on a contribution to ``Phases, Caustics and Chaos: Classical and Quantum Aspects'',}\\
\normalsize\emph{a meeting celebrating the science of Michael Berry, ICTS-TIFR, Bengaluru, October 2026}}
\author{Spenta R. Wadia\\[0.3em]
\small International Centre for Theoretical Sciences (ICTS-TIFR),\\
\small Tata Institute of Fundamental Research,\\
\small Shivakote, Bengaluru 560089, India\\[0.2em]
\small \texttt{spenta.wadia@icts.res.in}}
\date{}

\begin{document}
\maketitle

\begin{abstract}
We review the chiral anomaly from a historical perspective, from the $\pi^0\to\gamma\gamma$ puzzle and the work of Adler, Bell and Jackiw, to the Wess--Zumino--Witten (WZW) term. We then describe how the WZW term was motivated from quantum chromodynamics in 1983--85. Integrating out the non-Abelian gauge fields, that have a gapped spectrum, gives a Nambu--Jona-Lasinio type theory of quarks, and integrating out the quarks, which are gapped by chiral symmetry breaking, gives the WZW model. The fermion determinant splits into a regulator-dependent modulus, which gives the dynamics of mesons, and a universal phase, which is fixed by the anomaly and is the WZW term. Restricted to the collective coordinates of the baryon soliton, the WZW term imposes a $U(1)$ constraint that makes the baryon wave functions monopole harmonics and fixes their spin, statistics and flavour multiplets. We then explain that this phase is a Berry phase: the Schwinger terms of current algebra are the curvature of a $U(1)$ connection on the space of chiral field configurations, and the WZW term is its holonomy, which we call a Berry--Wilson loop. In $1+1$ dimensions we write the connection explicitly in terms of Euler angles, exhibiting a Dirac monopole whose quantised charge is the level of the current algebra.
\end{abstract}

\section{Introduction}

Anomalies are among the most robust facts of quantum field theory. A classical symmetry that cannot be preserved in the quantum theory leaves behind a precisely computable, one-loop exact, topologically quantised remnant, which survives all the way from the shortest to the longest distances. Berry's phase~\cite{Berry} is, in a sense as we shall explain, the same idea in quantum mechanics. When a system is transported adiabatically around a loop in the space of its parameters, its state need not return to itself, and the mismatch is a geometric phase determined by a connection and its curvature.

The purpose of this article, based on a talk at a meeting in honour of Michael Berry, is to describe how these two ideas meet. The one-line summary is:
\begin{quote}
\emph{A chiral anomaly is the statement that the fermion vacuum, transported around a loop in the space of background fields, does not come back to itself: a Berry phase.}
\end{quote}

The article is organised as follows. Section~\ref{sec:history} recalls the prelude: the $\pi^0$ puzzle, the Adler--Bell--Jackiw (ABJ) anomaly and its non-abelian generalisation by Bardeen. Section~\ref{sec:uses} describes some uses of anomalies in particle physics: anomaly cancellation in the Standard Model, the $U(1)$ problem, 't Hooft's anomaly matching, Fujikawa's derivation of the anomaly from the path-integral measure and its relation to the Atiyah--Singer index theorem, and Witten's global formulation of the Wess--Zumino term. Section~\ref{sec:QCD} reviews how the WZW term was motivated from QCD in 1983--85~\cite{DW,DSW}: integrating out the gauge fields, that have a gapped spectrum, gives a Nambu--Jona-Lasinio type theory of non-linear fermions, and integrating out the quarks, which are gapped by chiral symmetry breaking, gives the WZW model including the topological term. In Section~\ref{sec:baryons} we recall the quantisation of the baryon soliton and the resulting baryon wave functions which are monopole harmonics~\cite{JW}. Section~\ref{sec:berry} recalls Berry's phase and sets up a dictionary with the quark determinant. Section~\ref{sec:wilson} explains how the Schwinger terms of current algebra define a $U(1)$ connection on configuration space whose holonomy, the Berry--Wilson loop, is the WZW term~\cite{Wadia2021}, and writes the connection explicitly. Section~\ref{sec:outlook} summarises the main points, and Section~\ref{sec:notcovered} lists important topics that this article could not cover, among them gravitational anomalies, the Green--Schwarz mechanism, the mathematics of index theory, including the Witten index, and determinant line bundles, anomaly inflow and applications in condensed matter physics. Throughout, factors of $i$ and overall signs are convention dependent; we have tried to be consistent within each section.

\section{Prelude: the $\pi^0$ puzzle and the axial anomaly}
\label{sec:history}

\subsection{The puzzle}

In 1949 Steinberger computed the decay $\pi^0\to\gamma\gamma$ through a proton loop with a pseudoscalar pion--nucleon coupling, and obtained a rate surprisingly close to experiment~\cite{Steinberger}. Almost two decades later, Sutherland and Veltman~\cite{SV} showed that current algebra and the hypothesis of a partially conserved axial current (PCAC),
\begin{equation}
\partial_\mu j^\mu_{5,3} = f_\pi m_\pi^2\,\pi^0 ,
\end{equation}
together with gauge invariance, imply that the amplitude vanishes in the limit $m_\pi\to0$. Since the decay is observed with a width $\Gamma\approx7.8$~eV, either PCAC fails badly, or a step in the formal argument is wrong. The step that fails is the assumption that the classical relation $\partial_\mu j_5^\mu = 2im\,\bar\psi\gamma_5\psi$ holds as an operator identity in the quantum theory.

\subsection{Adler and Bell--Jackiw}

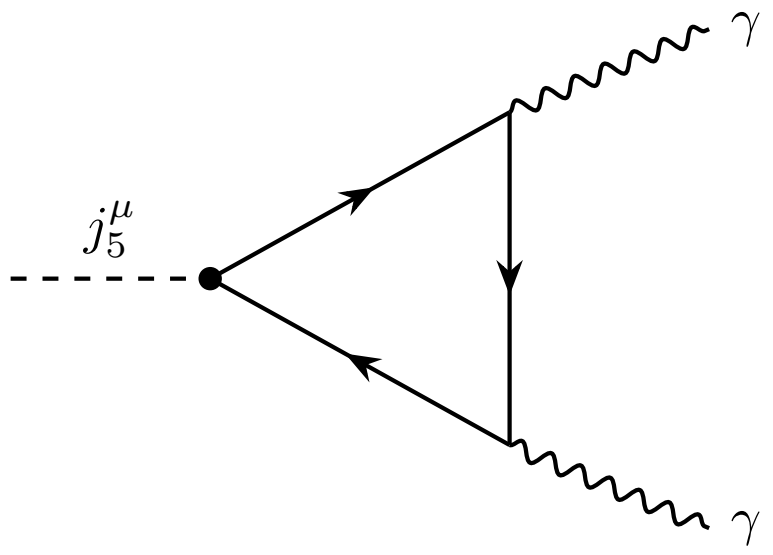
\begin{figure}[t]
\centering
\begin{tikzpicture}[scale=1.1,
  photon/.style={decorate,decoration={snake,amplitude=1.6pt,segment length=6pt}},
  fermion/.style={postaction={decorate},decoration={markings,mark=at position .55 with {\arrow{Stealth}}}}]
\coordinate (a) at (0,0);
\coordinate (b) at (1.8,1.0);
\coordinate (c) at (1.8,-1.0);
\draw[fermion,thick] (a)--(b);
\draw[fermion,thick] (b)--(c);
\draw[fermion,thick] (c)--(a);
\draw[thick,dashed] (-1.2,0)--(a) node[midway,above]{\small $j_5^\mu$};
\draw[photon,thick] (b)--(3.0,1.5) node[right]{\small $\gamma$};
\draw[photon,thick] (c)--(3.0,-1.5) node[right]{\small $\gamma$};
\fill (a) circle (2pt);
\end{tikzpicture}
\caption{The axial--vector--vector triangle graph responsible for the ABJ anomaly.}
\label{fig:triangle}
\end{figure}

Bell and Jackiw, in the $\sigma$-model with Pauli--Villars regularisation~\cite{BJ}, and Adler, in spinor electrodynamics~\cite{Adler}, showed that the triangle graph with one axial and two vector vertices (Fig.~\ref{fig:triangle}) is linearly divergent and depends on the routing of the loop momentum, so that no regulator can preserve both vector and axial current conservation. Keeping gauge invariance, one finds
\begin{equation}
\partial_\mu j_5^\mu = 2im\,\bar\psi\gamma_5\psi + \frac{e^2}{16\pi^2}\,\epsilon^{\mu\nu\rho\sigma}F_{\mu\nu}F_{\rho\sigma}.
\label{eq:ABJ}
\end{equation}
The extra term is a total derivative, $\epsilon^{\mu\nu\rho\sigma}F_{\mu\nu}F_{\rho\sigma} = \partial_\mu K^\mu$, yet it has physical consequences. It was in fact anticipated by Schwinger in 1951~\cite{Schwinger}: defining the axial current gauge invariantly, by point splitting with the phase $\exp\big(ie\int_x^{x+\epsilon}A\cdot dx\big)$, he found exactly this term.

\subsection{The decay rate and the number of colours}

The anomaly term gives the $\pi^0\to\gamma\gamma$ amplitude directly, and exactly in the chiral limit. With quark charges $Q_u=\tfrac23$, $Q_d=-\tfrac13$ and $N_c$ colours,
\begin{equation}
\Gamma(\pi^0\to\gamma\gamma) = \frac{\alpha^2 m_\pi^3}{64\pi^3 f_\pi^2}\Big[N_c\,(Q_u^2-Q_d^2)\Big]^2
= \frac{\alpha^2 m_\pi^3}{64\pi^3 f_\pi^2}\left(\frac{N_c}{3}\right)^2 .
\end{equation}
With $f_\pi\simeq92$~MeV this gives $\Gamma\simeq7.7$~eV for $N_c=3$, against $\approx0.86$~eV for $N_c=1$; equivalently a lifetime $\tau=\hbar/\Gamma\approx8.5\times10^{-17}$~s. The measured width, $\Gamma\approx7.8$~eV, was one of the early pieces of evidence that quarks come in three colours.

Adler and Bardeen showed that the coefficient of the anomaly is one-loop exact: it receives no corrections from higher orders of perturbation theory~\cite{AdlerBardeen}. This was an early hint that the anomaly is connected with topological aspects of quantum field theory.

\subsection{Bardeen's non-abelian anomaly}

Bardeen computed the general anomaly for fermions coupled to non-abelian vector and axial-vector fields, from the triangle, square and pentagon graphs~\cite{Bardeen}. In modern form, for a left-handed Weyl fermion in a gauge field $A=A_\mu^aT_a$,
\begin{equation}
(D_\mu J^\mu)_a = \frac{1}{24\pi^2}\,\epsilon^{\mu\nu\rho\sigma}\,\Tr\Big[T_a\,\partial_\mu\Big(A_\nu\partial_\rho A_\sigma + \tfrac12 A_\nu A_\rho A_\sigma\Big)\Big].
\label{eq:Bardeen}
\end{equation}
This anomaly satisfies the Wess--Zumino consistency conditions~\cite{WZ}, which express the fact that gauge transformations form a group. It is the anomaly that the effective action of the pions must reproduce, and that we recover from the quark determinant in Section~\ref{sec:QCD}.

\section{Some uses of anomalies in particle physics}
\label{sec:uses}

\subsection{Anomaly cancellation in the Standard Model}

A gauge anomaly would destroy gauge invariance, and with it unitarity and renormalisability, so it must cancel when summed over all fermions~\cite{GJ,BIM}. Each of the three quark--lepton generations, running around a loop coupled to the external gauge fields of the Standard Model, has quantum numbers such that all gauge anomalies cancel. This is crucial for the renormalisability of the Standard Model.

To see how this works, write the first generation ($u,d,\nu_e,e$) entirely in terms of left-handed Weyl fields: a right-handed field $f_R$ appears as its charge conjugate $f_R^{\,c}$, which is left-handed and carries opposite charges. With $Q=T_3+Y$, the quantum numbers and the anomaly coefficients are listed in Table~\ref{tab:SM}. Each row of the second table is a triangle graph with the indicated currents at its vertices, with coefficient $\propto\Tr[T^a\{T^b,T^c\}]$ summed over the left-handed fields; ``grav$^2U(1)_Y$'' denotes the mixed gravitational anomaly, with two graviton vertices and coefficient $\propto\sum Y$, whose cancellation is needed for consistency in curved spacetime. The last row is Witten's global $SU(2)$ anomaly~\cite{WittenSU2}, which requires an even number of doublets.

\begin{table}[t]
\centering
\begin{minipage}{0.37\textwidth}
\centering
\renewcommand{\arraystretch}{1.2}
\begin{tabular}{lccr}
\toprule
 & $SU(3)$ & $SU(2)$ & $Y$\\
\midrule
$(u,d)_L$ & $\mathbf 3$ & $\mathbf 2$ & $\tfrac16$\\
$u_R^{\,c}$ & $\bar{\mathbf 3}$ & $\mathbf 1$ & $-\tfrac23$\\
$d_R^{\,c}$ & $\bar{\mathbf 3}$ & $\mathbf 1$ & $\tfrac13$\\
\midrule
$(\nu_e,e)_L$ & $\mathbf 1$ & $\mathbf 2$ & $-\tfrac12$\\
$e_R^{\,c}$ & $\mathbf 1$ & $\mathbf 1$ & $1$\\
\bottomrule
\end{tabular}
\end{minipage}
\hfill
\begin{minipage}{0.58\textwidth}
\centering
\renewcommand{\arraystretch}{1.2}
\begin{tabular}{lccc}
\toprule
anomaly & quarks & leptons & total\\
\midrule
$SU(3)^3$ & $0$ & -- & $0$\\
$SU(3)^2\,U(1)_Y$ & $0$ & -- & $0$\\
grav$^2\,U(1)_Y$ & $0$ & $0$ & $0$\\
$SU(2)^2\,U(1)_Y$ & $+\tfrac12$ & $-\tfrac12$ & $0$\\
$U(1)_Y^3$ & $-\tfrac34$ & $+\tfrac34$ & $0$\\
\# $SU(2)$ doublets & $3$ & $1$ & $4$ (even)\\
\bottomrule
\end{tabular}
\end{minipage}
\caption{Left: one generation of the Standard Model as left-handed Weyl fields. Right: the gauge anomalies, separated into quark and lepton contributions.}
\label{tab:SM}
\end{table}

Quarks alone and leptons alone are each inconsistent; the anomalies cancel only between them. Each quark entry carries a factor $N_c$ (for example $N_c\cdot\tfrac16$ against $-\tfrac12$ in the $SU(2)^2U(1)_Y$ row), so cancellation requires $N_c=3$. A generation cannot be incomplete: once the $\tau$ lepton (1975) and the $b$ quark (1977) were found, the third generation needed a partner, the top quark, finally discovered at Fermilab in 1995 by the CDF and D\O\ collaborations, with a mass of about $173$~GeV. The second and third generations repeat the pattern of the first.

\subsection{Instantons, the $U(1)$ problem and anomaly matching}

In QCD the divergence of the flavour-singlet axial current is proportional to $\Tr\,G\tilde G$, and there exist (euclidean) gauge field configurations, instantons, with integer topological charge $\nu=\frac{1}{32\pi^2}\int\Tr\,G\tilde G$. 't Hooft~\cite{tHooft76} showed that the associated fermion zero modes lead to $\Delta Q_5 = 2N_f\nu$: the axial $U(1)$ is not a symmetry of QCD, and there is no ninth Goldstone boson.

't Hooft also formulated the principle of anomaly matching~\cite{tHooft79}. Weakly gauge a global symmetry, adding spectator fermions to cancel its anomaly. Since the gauged theory must be consistent at all scales, the anomaly of the original degrees of freedom must be reproduced at every lower energy scale, in order to cancel the anomaly of the weakly coupled spectators. 't Hooft's argument thus shows that anomalies are invariant under the renormalisation group flow: the anomaly computed in the ultraviolet equals the anomaly of the infrared degrees of freedom. In QCD with spontaneously broken chiral symmetry, the infrared degrees of freedom are the Nambu-Goldstone bosons, and the anomaly is reproduced by the Wess--Zumino term.

\subsection{Fujikawa: the anomaly lives in the measure}

Fujikawa~\cite{Fujikawa} showed that the anomaly arises from the non-invariance of the fermion measure in the path integral. Consider Euclidean QED and expand the fermion fields in eigenmodes of the Dirac operator, $\slashed D\phi_n=\lambda_n\phi_n$. Under a local chiral rotation $\psi\to e^{i\alpha(x)\gamma_5}\psi$, $\bar\psi\to\bar\psi e^{i\alpha(x)\gamma_5}$, the measure changes as
\begin{equation}
\mathcal D\psi\,\mathcal D\bar\psi\;\to\;\exp\Big(-2i\!\int d^4x\,\alpha(x)\,\mathcal A(x)\Big)\,\mathcal D\psi\,\mathcal D\bar\psi,\qquad
\mathcal A(x) = \sum_n\phi_n^\dagger(x)\,\gamma_5\,\phi_n(x).
\end{equation}
As it stands $\mathcal A(x)$ is ill defined, being of the form $\Tr\gamma_5\times\delta(0)$, and must be regularised in a gauge-invariant way:
\begin{equation}
\mathcal A(x) = \lim_{M\to\infty}\Tr\big[\gamma_5\,e^{-\slashed D^2/M^2}\big](x,x).
\end{equation}
The heat kernel is evaluated using $\slashed D^2 = D^\mu D_\mu + \frac{e}{2}\sigma^{\mu\nu}F_{\mu\nu}$ and expanding in $\sigma F/M^2$. The trace with $\gamma_5$ needs exactly two factors of $\sigma F$; the momentum integral gives $M^4/16\pi^2$, which cancels the $1/M^4$ from the expansion. Terms with fewer factors of $F$ vanish under the trace, and terms with more are suppressed by powers of $1/M^2$. The result is
\begin{equation}
\mathcal A(x) = -\frac{e^2}{32\pi^2}\,\epsilon^{\mu\nu\rho\sigma}F_{\mu\nu}F_{\rho\sigma},
\end{equation}
with a convention-dependent sign, which reproduces the ABJ equation \eqref{eq:ABJ}.

\subsection{The Atiyah--Singer index theorem}

The topological character of the anomaly is made manifest by taking $\alpha$ constant, so that the Jacobian involves $\int d^4x\,\mathcal A(x)$. Since $\{\gamma_5,\slashed D\}=0$, if $\slashed D\phi_n=\lambda_n\phi_n$ then $\slashed D(\gamma_5\phi_n)=-\lambda_n\gamma_5\phi_n$. For $\lambda_n\neq0$ the modes $\phi_n$ and $\gamma_5\phi_n$ have different eigenvalues and are therefore orthogonal, so that
\begin{equation}
\int d^4x\;\phi_n^\dagger\gamma_5\phi_n = 0\qquad(\lambda_n\neq0).
\end{equation}
The non-zero modes pair up and cancel; only the zero modes survive, and they can be chosen with definite chirality $\gamma_5=\pm1$. The regulator drops out, and
\begin{equation}
\int d^4x\;\mathcal A = n_+ - n_- = \operatorname{ind}\slashed D
= \frac{e^2}{32\pi^2}\int d^4x\,\epsilon^{\mu\nu\rho\sigma}F_{\mu\nu}F_{\rho\sigma}
= \frac{1}{8\pi^2}\int (eF)\wedge(eF)\;\in\;\mathbb Z .
\end{equation}
This is the Atiyah--Singer index theorem~\cite{AS}: the analytic index equals a topological invariant. It explains three facts at once: the anomaly coefficient is one-loop exact, since an integer cannot receive small corrections; instantons with $\nu\neq0$ force fermion zero modes and the 't Hooft vertex $\Delta Q_5 = 2N_f\nu$; and the anomaly density is the local index density.

\subsection{Witten's global formulation of the Wess--Zumino term}

Wess and Zumino~\cite{WZ} constructed the effective action of the pseudoscalar mesons that reproduces the anomaly. Witten~\cite{Witten83} gave it a global, geometric formulation. The chiral Lagrangian $\frac{f_\pi^2}{4}\Tr\,\partial_\mu U\partial^\mu U^\dagger$ has spurious discrete symmetries that forbid, for example, the decay $K^+K^-\to\pi^+\pi^-\pi^0$; the Wess--Zumino term breaks them. It cannot be written as a local four-dimensional Lagrangian. Instead one extends $U$ to a five-dimensional disc $D^5$ whose boundary is (compactified) spacetime, $\partial D^5=S^4$:
\begin{equation}
\Gamma_{WZW} = -\frac{i}{240\pi^2}\int_{D^5}d^5x\;\epsilon^{ijklm}\,\Tr\big[L_iL_jL_kL_lL_m\big],\qquad L_i = U^\dagger\partial_iU.
\label{eq:WZW}
\end{equation}
Two extensions differ by an integral over $S^5$, and since $\pi_5(SU(3))=\mathbb Z$ this is $2\pi$ times an integer. For $e^{iN\Gamma}$ to be well defined, $N$ must be an integer. Gauging the WZW term reproduces $\pi^0\to\gamma\gamma$ and fixes $N=N_c$; and Skyrmions are fermions when $N_c$ is odd~\cite{Witten83}. Witten drew the analogy with a charged particle moving on a sphere around a Dirac monopole, whose action is defined only via a two-dimensional disc and is single valued only if the product of electric and magnetic charges is quantised. This picture will reappear in Sections~\ref{sec:berry} and~\ref{sec:wilson}.

\section{From QCD to the WZW term}
\label{sec:QCD}

In this section we review how the WZW term, including its quantised coefficient, arises from QCD~\cite{DW,DSW}. Related work around the same time includes~\cite{GW,BMNT,BNRS,DHF}.

\subsection{A theory of non-linear fermions}

One fundamental fact about QCD that is exhibited in lattice gauge theory calculations, is that the pure gauge sector has a mass gap. To study chiral symmetries of QCD in the conceptual framework of the Kadanoff--Wilson renormalisation group, we imagine integrating over the short-wavelength degrees of freedom of the gauge field, up to length scales of order the inverse glueball mass. Since the gauge field then has a finite correlation length, the resulting theory only evaluates correlations of gauge-invariant fermionic operators, much as Fermi's theory of weak interactions emerges from the electroweak theory at wavelengths much longer than $1/M_W$. The result is a theory of non-linear fermions that generalises the Nambu--Jona-Lasinio model~\cite{NJL}:
\begin{align}
-i\mathcal L = {}& \bar\psi_{L\alpha j}\,i\slashed\partial\,\psi_{L\alpha j} + \bar\psi_{R\alpha j}\,i\slashed\partial\,\psi_{R\alpha j}
+ g_1^2\,\big(\bar\psi_{L\alpha j}\psi_{R\alpha k}\big)\big(\bar\psi_{R\beta k}\psi_{L\beta j}\big)\nonumber\\
&- i\,\frac{g_2^2}{4}\Big[\big(\bar\psi_{L\alpha j}\gamma_\mu\psi_{L\alpha k}\big)\big(\bar\psi_{L\beta k}\gamma_\mu\psi_{L\beta j}\big) + L\leftrightarrow R\Big] + \cdots,
\label{eq:NJL}
\end{align}
where $\alpha,\beta=1,\dots,N$ are colour and $j,k=1,\dots,n$ flavour indices, $g_1,g_2$ have mass dimension $-1$, and the cutoff $\Lambda$ is below the glueball mass. The theory has a global $U(n)\times U(n)$ chiral symmetry.

Introducing auxiliary gauge-invariant collective fields, a complex flavour matrix $M$ and vector fields $L_\mu$, $R_\mu$ coupling to the left- and right-handed quarks, we can rewrite \eqref{eq:NJL} as
\begin{align}
-i\mathcal L = {}& \bar\psi_L\,i\slashed\partial\,\psi_L + \bar\psi_R\,i\slashed\partial\,\psi_R + \bar\psi_L M\psi_R + \bar\psi_R M^\dagger\psi_L
+ i\big(\bar\psi_L\slashed L\psi_L + \bar\psi_R\slashed R\psi_R\big)\nonumber\\
&+ \frac{i}{g_1^2}\,\tr\,M^\dagger M + \frac{i}{g_2^2}\,\tr\big(L_\mu^2 + R_\mu^2\big);
\end{align}
the Gaussian integral over $M$, $L_\mu$, $R_\mu$ gives back \eqref{eq:NJL}. Under $(V_1,V_2)\in U(n)\times U(n)$, $\psi_L\to V_1\psi_L$, $\psi_R\to V_2\psi_R$, $M\to V_1MV_2^\dagger$, $L_\mu\to V_1L_\mu V_1^\dagger$ and $R_\mu\to V_2R_\mu V_2^\dagger$.

\subsection{Quarks in a slowly varying chiral background}

In this background the quarks see the Dirac operator
\begin{equation}
D = i\slashed\partial - \slashed L P_L - \slashed R P_R + iMP_L + iM^\dagger P_R .
\end{equation}
$P_L$ and $P_R$ are chiral projection operators. At large $N$, for $\Lambda^2g_1^2 > 8\pi^2/N$, chiral symmetry is spontaneously broken: $M=\bar H\,U$, with $\bar H\sim\langle\bar\psi\psi\rangle$ a constituent quark mass and $U(x)$ the pion field. The quarks are therefore gapped. Integrating them out and expanding in long wavelengths, $k\ll\bar H$, everything is contained in the fermion determinant, which we split into its modulus and phase:
\begin{equation}
N\ln\det D = N\ln|\det D| + iN\Delta .
\label{eq:modphase}
\end{equation}
The determinant is computed by the heat-kernel (Schwinger proper-time) method,
\begin{equation}
\ln\det D = -\frac12\int_{1/\Lambda^2}^\infty\frac{ds}{s}\,\Tr\,e^{-s\hat D^2},\qquad \hat D = i\gamma_5D,
\end{equation}
expanded in derivatives of the background fields. The choice of $\hat D$ ensures that the vector currents are conserved, as QCD requires~\cite{VW}, and the anomaly agrees with Bardeen's result.

\subsection{Two pieces of the determinant, two kinds of physics}

The \emph{modulus} $\ln|\det D|$ gives the dynamics of the mesons: the pion kinetic term, four-derivative terms, and the couplings and masses of vector and axial-vector mesons, for example $m_A^2-m_V^2=6\bar H^2$~\cite{DSW}. At long wavelengths the pion action is
\begin{equation}
S = \int d^4x\,\Big[\tfrac12F_\pi^2\,\tr(L^\mu L_\mu) + \frac{N(d_4-d_6)}{16}\,\tr\,[L_\mu,L_\nu]^2 + \frac{Nd_5}{16}\,\tr\,(L_\mu L^\mu)^2\Big] + N\,\Gamma_{WZW} + \cdots,
\label{eq:Seff}
\end{equation}
with $L_\mu = i(\partial_\mu U)U^\dagger$. The first four-derivative term is the Skyrme term; the second is an additional term, so that the result is a generalised Skyrme model with soliton solutions. The coefficients $d_i$ are computed from the quark loop as functions of $\bar H^2/\Lambda^2$; they depend on the cutoff and on the choice of regulator, and are therefore \emph{not universal}.

The \emph{phase} $\Delta$, by contrast, gives the WZW term $N\Gamma_{WZW}$ with a coefficient independent of $\bar H$, of the cutoff and of the regulator. It is \emph{universal}, fixed entirely by the anomaly, and carries baryon number and all the anomalous processes. As we shall see in Section~\ref{sec:berry}, this division between a regulator-dependent modulus and a universal phase is similar to Berry's division between a dynamical and a geometric phase.

\subsection{The phase is determined by the anomaly}

The phase is not invariant under \emph{local} chiral rotations, because the fermion measure is not~\cite{Fujikawa}. Since our regularisation conserves the vector currents, the variation of the phase agrees with Bardeen's anomaly \eqref{eq:Bardeen}. Restricted to the pion field,
\begin{equation}
G_a\,(i\Gamma) = \frac{1}{48\pi^2}\,\epsilon^{\mu\nu\rho\sigma}\,\tr\big(t_a\,L_\mu L_\nu L_\rho L_\sigma\big),
\label{eq:WZeq}
\end{equation}
where $G_a$ generates $U(n)$ rotations on the space of configurations $U(x)$~\cite{DW}. As Witten emphasised, this equation cannot be integrated to a local four-dimensional action. But the Wess--Zumino consistency condition is a zero-curvature condition on configuration space~\cite{DSW}, so the equation can be integrated along any path in configuration space, with a path-independent answer. Here configuration space is the Riemannian manifold of field configurations $U(x)$, with metric
\begin{equation}
ds^2 = \int d^4x\,\tr\big(dU\,dU^\dagger\big).
\end{equation}

\subsection{Integrating the anomaly along a path in configuration space}

Take a path $U(x,s)$, $s\in[0,1]$, from $U=1$ to the configuration of interest, with $\Gamma(0)=0$. Along it, with $L_s = i(\partial_sU)U^\dagger$, equation~\eqref{eq:WZeq} becomes
\begin{equation}
\frac{d\Gamma}{ds} = \frac{1}{48\pi^2}\int d^4x\;\epsilon^{\mu\nu\rho\sigma}\,\tr\big(L_s\,L_\mu L_\nu L_\rho L_\sigma\big).
\label{eq:dGamma}
\end{equation}
The coordinates $(s,x)$ sweep out a five-dimensional disc $D^5$ bounded by spacetime $S^4$, and the result is Witten's formula~\cite{DW},
\begin{equation}
\Gamma = -\frac{1}{240\pi^2}\int_{D^5}d\Sigma^{ijklm}\,\tr\big(L_iL_jL_kL_lL_m\big).
\end{equation}
The integrand is closed, so the answer is independent of the path; $\pi_5(SU(3))=\mathbb Z$ makes $e^{iN\Gamma}$ single valued only for integer $N$, and here $N$ is the number of colours. Since real representations are anomaly free, there is no WZW term for two flavours. The WZW term is thus \emph{derived} from the quarks rather than postulated.

\subsection{What the phase carries}

\paragraph{Baryon number.} The baryon current is
\begin{equation}
B_\mu = -\frac{i}{N}\,\tr\Big(\frac{\delta}{\delta L_\mu}+\frac{\delta}{\delta R_\mu}\Big)\ln\det D .
\end{equation}
The modulus is invariant under vector gauge transformations, so only the phase contributes. At long wavelengths~\cite{DW,DSW}
\begin{equation}
B_\mu = -\frac{1}{24\pi^2}\,\epsilon_{\mu\nu\rho\sigma}\,\tr\big(\partial^\nu U U^{-1}\partial^\rho UU^{-1}\partial^\sigma UU^{-1}\big),
\end{equation}
a topological current: baryon number equals winding number~\cite{GW,BNRS}, and baryons are the solitons of \eqref{eq:Seff}.

\paragraph{Anomalous processes.} Gauging the quark charge $Q$ in the fermion theory automatically gives the gauged WZW term, and hence $\pi^0\to\gamma\gamma$ and the other current-algebra anomalies, in agreement with Witten~\cite{Witten83}.

\paragraph{A modified Gauss law.} The anomaly term modifies the Gauss-law constraint on configuration space~\cite{DSW}, a first sign that the phase acts like a connection.

\medskip
In short, everything about the quarks that survives at long distances with a universal coefficient sits in the phase of the determinant.

\section{Collective coordinates and baryon quantum numbers}
\label{sec:baryons}

Restricted to the collective coordinates of the soliton, the phase fixes the spin, statistics and flavour multiplets of the baryons~\cite{JW}; see also~\cite{Guadagnini,MNP}. Consider the action
\begin{equation}
S(U) = N\big[S_0(U) + \Gamma(U)\big],
\end{equation}
where $S_0$ is the Skyrme action and $\Gamma$ the WZW term. At large $N$, $1/N$ plays the role of $\hbar$, a WKB parameter, and the soliton is quantised by the collective-coordinate method of Gervais and Sakita~\cite{GS}, which goes back to Wentzel's strong-coupling theory~\cite{Wentzel}. Write
\begin{equation}
U(t,\mathbf x) = \Omega(t)\,U_0(\mathbf x)\,\Omega^{-1}(t),
\end{equation}
where $U_0$ is the hedgehog solution, with baryon number one, embedded in $SU(3)$, and $\Omega(t)\in SU(3)$. Since $U_0$ commutes with the hypercharge $Y=\frac13\mathrm{diag}(1,1,-2)$, $\Omega$ and $\Omega h$ with $h=e^{i\epsilon Y}$ describe the same configuration: the collective coordinate space is $SU(3)/U(1)$, which we treat as $SU(3)$ with a local $U(1)$ symmetry.

Substituting this $U$ into $S(U)$ gives the collective-coordinate Lagrangian
\begin{equation}
L = -\tfrac12 I_{cd}\,\Tr(t_c\omega_R)\,\Tr(t_d\omega_R) - \frac{N}{48\pi^2}\,\Tr(\omega_RK),\qquad \omega_R = \Omega^{-1}\dot\Omega,
\label{eq:Lcoll}
\end{equation}
where $I_{cd}$ is the moment-of-inertia tensor and $K\propto\mathrm{diag}(1,1,0)$ is built from the baryon density of the hedgehog. The second term comes from the WZW term. It is linear in velocities and, as noted in~\cite{JW}, ``imposes a crucial boundary condition on the baryon wave functions'': the $U(1)$ gauge condition
\begin{equation}
Y_R = \frac{N}{3}
\label{eq:YR}
\end{equation}
on the right hypercharge of the wave functions.

\subsection{Baryon quantum numbers}

Because the moment-of-inertia tensor has a zero eigenvalue in the hypercharge direction, the system is constrained. In the Hamiltonian formalism the constraint reads $Y_R=N/3$, which, in the words of~\cite{JW}, is ``like the Gauss constraint on physical states in gauge theories'': it arises from the local $U(1)$ symmetry introduced by describing the seven-dimensional space $SU(3)/U(1)$ with eight variables. Physical wave functions satisfy
\begin{equation}
\hat Y_R\,\psi(\Omega) = \frac{N}{3}\,\psi(\Omega),\qquad e^{i\alpha\hat Y_R}\psi(\Omega) = \psi(\Omega\,e^{i\alpha Y}).
\end{equation}
Since $e^{6\pi iY}=1$, these imply $\psi(\Omega)=e^{2\pi iN}\psi(\Omega)$, and single-valuedness requires $N\in\mathbb Z$.

The system is a generalised symmetric top: the space of variables is $SU(3)$, and $\Omega\to V\Omega W$ is a symmetry with $V\in SU(3)$ (flavour) and $W$ in the $SU(2)\times U(1)$ subgroup acting on the right, whose $SU(2)$ part is identified with spatial rotations. The wave functions are matrix elements $D^{pq}_{\alpha\beta}(\Omega)$ of the $SU(3)$ representations $D(p,q)$, where $\beta=(I,I_3,Y)$ labels flavour and $\alpha=(I',I_3',Y')$ labels spin and right hypercharge. Without the WZW term only $Y'=0$ would be allowed. With it, only states with $Y'=N/3$ survive: the wave functions are not functions on $SU(3)/U(1)$ but sections of a line bundle of charge $N/3$ over it, that is, \emph{monopole harmonics}~\cite{WuYang}, the analogue of $D^j_{m,q}$ for a charge in a monopole field with $q=eg$. Representation theory of $SU(3)$ then gives~\cite{JW}
\begin{equation}
p-q = N+3l,\qquad I_3' = \frac N2 + k,\qquad l,k\in\mathbb Z .
\end{equation}
The first condition says that the triality of the allowed flavour representations equals $N$ modulo 3; the second that the baryons are fermions for odd $N$ and bosons for even $N$. For $N=3$ only triality-zero representations occur: the lowest are the flavour octet, in which $Y'=1$ forces spin $\tfrac12$, and the decuplet, with spin $\tfrac32$. These are exactly the representations in which baryons are observed.

The Hamiltonian is a sum of quadratic Casimirs. With $C_1$, $C_2$ the two independent moments of inertia and $C_{pq}$ the quadratic Casimir of $D(p,q)$, the energies are~\cite{JW}
\begin{equation}
E(p,q;I') = \Big(\frac1{C_1}-\frac1{C_2}\Big)I'(I'+1) + \frac1{C_2}\Big(C_{pq}-\frac{N^2}{12}\Big),
\end{equation}
which gives for the nucleon and the $\Delta$
\begin{equation}
M_N = M_{\rm cl} + \frac34\Big(\frac1{C_1}+\frac2{C_2}\Big),\qquad
M_\Delta = M_{\rm cl} + \frac34\Big(\frac5{C_1}+\frac2{C_2}\Big),
\end{equation}
so that $M_\Delta-M_N = 3/C_1$. Here $M_{\rm cl}$ is of order $N$ and $1/C_{1,2}$ of order $1/N$. The division of labour is instructive: the geometric term fixes the quantum numbers, while the dynamical part, through $C_1$ and $C_2$, fixes the masses. As emphasised in~\cite{JW}, the WZW term ``carries `global' and `length scale independent' information'', and is independent of the density field $H$ that describes the core of the baryon.

This generalises Wentzel's collective-coordinate model of the pion--nucleon system~\cite{Wentzel}, whose internal symmetry group was isospin $SU(2)$, to flavour $SU(3)$ with the WZW term.

In the language of the next section, the WZW term in \eqref{eq:Lcoll} is a Berry connection on the space of collective coordinates. Writing $\omega_R = i\omega^at_a$, it is $(N/\sqrt6)\,\omega^8$, and its curvature is proportional to $f_{8bc}\,\omega^b\wedge\omega^c$, non-zero only for $b,c\in\{4,5,6,7\}$: it lives on $SU(3)/U(2)\cong\mathbb{CP}^2$. The topological charge of the soliton in space thus becomes a Berry flux in the space of its collective coordinates.

\section{The Berry phase connection}
\label{sec:berry}

\subsection{Berry's geometric phase}

Consider a Hamiltonian $H(\mathbf R)$ depending on slowly varying parameters $\mathbf R(t)$, and a non-degenerate level $|n,\mathbf R\rangle$. Under adiabatic evolution around a closed loop $C$ in parameter space~\cite{Berry},
\begin{equation}
|\psi(T)\rangle = e^{i\gamma_n(C)}\,e^{-\frac{i}{\hbar}\int_0^TE_n\,dt}\,|\psi(0)\rangle,\qquad
\gamma_n(C) = \oint_C\mathcal A_n,\qquad \mathcal A_n = i\langle n|\nabla_{\mathbf R}\,n\rangle\cdot d\mathbf R .
\end{equation}
$\mathcal A_n$ is a $U(1)$ connection on parameter space, and a change of phase $|n\rangle\to e^{i\chi(\mathbf R)}|n\rangle$ is a gauge transformation. The Berry phase $\gamma_n$ is gauge invariant modulo $2\pi$, and by Stokes' theorem equals the flux of the curvature $\mathcal F_n = d\mathcal A_n$ through any surface bounded by $C$. Simon identified it as the holonomy of a line bundle~\cite{Simon}, and Wilczek and Zee gave the non-abelian generalisation for degenerate levels~\cite{WilczekZee}.

\subsection{The canonical example: a spin in a magnetic field}

Let $H=-\mu\,\mathbf B\cdot\mathbf S$ for a spin $s$. Both of the pictures below use one family of states, the spin coherent states $|\hat{\mathbf n}\rangle$ pointing along the unit vector $\hat{\mathbf n}$, with $\langle\hat{\mathbf n}|\mathbf S|\hat{\mathbf n}\rangle = s\,\hat{\mathbf n}$. The key identity is that for any closed loop $\hat{\mathbf n}(t)$ on $S^2$,
\begin{equation}
i\oint\langle\hat{\mathbf n}|\tfrac{d}{dt}|\hat{\mathbf n}\rangle\,dt = -s\,\Omega,
\label{eq:spinBerry}
\end{equation}
where $\Omega$ is the solid angle enclosed. The curvature is that of a monopole of strength $s$; since $\Omega$ is defined only up to $\Omega\to\Omega-4\pi$, single-valuedness requires $2s\in\mathbb Z$.

\emph{Berry's picture.} The field $\mathbf B(t)$ is rotated around a loop slowly compared with the Larmor frequency $\mu|\mathbf B|$. A spin that starts in its ground state stays aligned, $\hat{\mathbf n}(t)=\hat{\mathbf b}(t)=\mathbf B/|\mathbf B|$, and acquires a dynamical phase plus the geometric phase $-s\,\Omega$, with $\Omega$ the solid angle swept by $\hat{\mathbf b}$. In $\mathbf B$-space the monopole sits at $\mathbf B=0$, where the $2s+1$ levels cross.

\emph{The path integral.} Here $\hat{\mathbf n}(t)$ is the integration variable, and each path is weighted by $e^{iS}$ with
\begin{equation}
S[\hat{\mathbf n}] = \int i\langle\hat{\mathbf n}|\partial_t\hat{\mathbf n}\rangle\,dt + \int\mu s\,\mathbf B\cdot\hat{\mathbf n}\,dt .
\end{equation}
The first term is a Wess--Zumino term in $0+1$ dimensions, which on a disc reads $s\int_{D^2}\hat{\mathbf n}\cdot(\partial_t\hat{\mathbf n}\times\partial_\tau\hat{\mathbf n})$: it is the Berry phase of each path. It is this term that makes $\hat{\mathbf n}$ a quantum spin, with $2s+1$ states and half-integer $s$ allowed. The classical path is Larmor precession.

The two pictures are connected as follows: Berry's slowly rotated field selects one path in the sum, $\hat{\mathbf n}(t)=\hat{\mathbf b}(t)$, and on that path the Wess--Zumino term is Berry's phase $-s\,\Omega$. The general lesson~\cite{Stone,Aitchison,Jackiw} is that when fast degrees of freedom are integrated out in the adiabatic (Born--Oppenheimer) approximation, the effective action of the slow variables acquires a term linear in velocities, $i\langle0|\partial_t|0\rangle$, which is the Berry connection; when its curvature is that of a monopole it cannot be written globally as a local term and is a Wess--Zumino term, with a quantised coefficient.

\subsection{A dictionary}

The quark determinant of Section~\ref{sec:QCD} fits this pattern exactly. The fast degrees of freedom are the quarks of the Dirac sea, gapped by $\bar H$; the slow variables are the chiral field $U(\mathbf x,t)$; and the WZW term is the geometric phase of the fermion vacuum $|\Omega[U]\rangle$ as $U$ varies. Table~\ref{tab:dict} summarises the correspondence.

\begin{table}[h]
\centering
\renewcommand{\arraystretch}{1.3}
\begin{tabular}{>{\raggedright\arraybackslash}p{0.42\textwidth}>{\raggedright\arraybackslash}p{0.48\textwidth}}
\toprule
\textbf{Berry's adiabatic phase} & \textbf{Quarks in a chiral background} \\
\midrule
$H(\mathbf R)$, slowly varying parameters $\mathbf R(t)$ & Dirac operator $D[U]$, slowly varying $U(\mathbf x,t)$ \\
Parameter space of $\mathbf R$ & Configuration space $\mathcal M=\mathrm{Map}(S^3,SU(N_f))$ \\
Non-degenerate level $|n,\mathbf R\rangle$ & Fermion vacuum $|\Omega[U]\rangle$, gapped by $\bar H$ \\
Connection $\mathcal A_n = i\langle n|\nabla_{\mathbf R}\,n\rangle$ & $U(1)$ connection on $\mathcal M$ \\
Curvature $\mathcal F_n = d\mathcal A_n$ & Schwinger term \\
Berry phase $\gamma_n=\oint_C\mathcal A_n=\iint\mathcal F_n$ & Phase of $\det D$: the WZW term on $D^5$ \\
Dynamical phase $\int E_n\,dt$ & Modulus $\ln|\det D|$ (regulator dependent) \\
Gauge invariant mod $2\pi$; quantised flux & $\pi_5(SU(3))=\mathbb Z$; $N_c\in\mathbb Z$ \\
\bottomrule
\end{tabular}
\caption{Berry's phase and the quark determinant.}
\label{tab:dict}
\end{table}

Concretely, consider equation \eqref{eq:dGamma} of Section~\ref{sec:QCD}, which is the equation integrated by Dhar and Wadia~\cite{DW}, who chose spacetime to be $S^3\times S^1$ with time periodic. A static configuration $U(\mathbf x)$ is then a point of $\mathcal M=\mathrm{Map}(S^3,SU(N_f))$, and the time evolution $U(\mathbf x,t)$ is a closed loop in $\mathcal M$. In \eqref{eq:dGamma}, besides $L_s$, exactly one of the four indices of the $\epsilon$-tensor is temporal, so that
\begin{equation}
\Gamma = \int_0^1\! ds\oint dt\;\mathcal F_{st},\qquad
\mathcal F_{st}\propto\int_{S^3}d^3x\;\epsilon^{ijk}\,\tr\big(L_s\,L_t\,L_iL_jL_k\big):
\end{equation}
$\Gamma$ is the flux, through the surface swept out by the family of loops $s\mapsto U(\mathbf x,t;s)$, of a two-form $\mathcal F$ on $\mathcal M$ obtained by integrating the five-form $\tr(dU\,U^{-1})^5$ over $S^3$. This is Berry's formula $\gamma=\iint\mathcal F$. Witten's five-dimensional disc is the surface swept out in $\mathcal M$, and the quantisation of $N$ follows from $\pi_2(\mathcal M)=\pi_5(SU(3))=\mathbb Z$: a monopole in field space. The Dhar--Wadia derivation~\cite{DW}, submitted in September 1983, a few months before Berry's paper appeared in March 1984, is thus an early field-theoretic computation of a many-body Berry phase, made before the language existed; the explicit link between Wess--Zumino terms and Berry phases was drawn later~\cite{Stone,Aitchison,NiemiSemenoff,Jackiw}.

\section{The WZW term as a Berry--Wilson loop in configuration space}
\label{sec:wilson}

The connection on configuration space can be made completely explicit~\cite{Wadia2021}. We work in $1+1$ dimensions with $G=SU(2)$; the construction extends to $SU(N)$ and to $3+1$ dimensions.

\subsection{Current algebra, Schwinger terms and a $U(1)$ gauge connection}

Let the configuration space be $\mathcal M=\mathrm{Map}(S^1,G)$, $x\mapsto U(x)$. The current algebra with Schwinger terms reads
\begin{equation}
[L_0^\alpha,L_0^\beta] = i\lambda^2 f_{\alpha\beta\gamma}L_0^\gamma + i\alpha f_{\alpha\beta\gamma}L_1^\gamma,\qquad
[L_0^\alpha(x),L_1^\beta(y)] = i\lambda^2 f_{\alpha\beta\gamma}L_1^\gamma + \tfrac{\lambda^4}{2}\,\delta^{\alpha\beta}\partial_x\delta(x-y),
\end{equation}
with $4\pi\alpha\lambda^{-2}=k\in\mathbb Z$ (local factors of $\delta(x-y)$ are suppressed). We realise the currents in terms of a $U(1)$ connection $A^\alpha(U(x))$ on $\mathcal M$:
\begin{equation}
l_0^\alpha(x) = \pi^\alpha(x) - A^\alpha,\qquad \pi^\alpha = \tr\Big(t^\alpha U\frac{\partial}{\partial U}\Big),\qquad
l_1^\alpha = \tr\big(t^\alpha\,\partial_xU\,U^{-1}\big),
\end{equation}
with gauge transformations $A\to A+\delta\phi$. With $\delta_{\alpha x}U(y)=t^\alpha U(y)\,\delta(x-y)$, the gauge-invariant field strength is
\begin{equation}
F_{\alpha x,\beta y} = i\,\delta_{\alpha x}A_{\beta y} - i\,\delta_{\beta y}A_{\alpha x} - f_{\alpha\beta\gamma}A_{\gamma x}\,\delta(x-y),
\end{equation}
and the covariant derivatives $l_0^\alpha$ do not commute:
\begin{equation}
[l_0^\alpha(x),l_0^\beta(y)] = -if_{\alpha\beta\gamma}\,l_0^\gamma\,\delta(x-y) + i\,F_{\alpha x,\beta y}.
\end{equation}
The current algebra is reproduced if and only if
\begin{equation}
F_{\alpha x,\beta y} = \frac{k}{4\pi}\,f_{\alpha\beta\gamma}\,l_1^\gamma\,\delta(x-y).
\label{eq:Fk}
\end{equation}
That is, \emph{the Schwinger term is the curvature of a $U(1)$ gauge field on configuration space}: a Berry curvature.

\subsection{The Berry--Wilson loop}

The holonomy of this connection around a closed curve $t\mapsto U(x,t)$ in $\mathcal M$ is a Berry phase in Berry's language and a Wilson loop in the language of gauge theory; we call it a \emph{Berry--Wilson loop}. Using Stokes' theorem and \eqref{eq:Fk},
\begin{align}
\Phi &= \oint dt\!\int_0^{2\pi}\!dx\;A_\alpha\,l_t^\alpha
= \int_0^1\!ds\oint dt\!\int dx\;F_{\alpha\beta}\,l_t^\alpha l_s^\beta\nonumber\\
&= \frac{k}{12\pi}\int_D d^3x\,\epsilon^{ijk}\,\tr\big(\partial_iUU^{-1}\partial_jUU^{-1}\partial_kUU^{-1}\big),
\label{eq:Phi}
\end{align}
which is the WZW term, normalised so that the flux through any closed surface in $\mathcal M$ is $2\pi k$.\footnote{This corrects a factor of $2$ in the final expression of~\cite{Wadia2021}, where the coefficient $k/24\pi$ appears; the field strength \eqref{eq:Fk} and the explicit potential of Section~\ref{sec:euler} both correspond to $k/12\pi$.}

In the Schr\"odinger picture, the wave functional carries the phase
\begin{equation}
\Psi[\{U(x)\},T]\sim\exp\Big(i\int^Tdt\int dx\,A_a\,\dot\theta^a\Big),
\end{equation}
and around a closed curve it changes by $e^{i\Phi}$: the wave functional carries a projective representation of the current algebra, encoded in a topologically non-trivial $U(1)$ gauge field. The potential $A$ necessarily has a Dirac-like singularity, and $k$ is the quantised charge of a ``magnetic monopole'' on $\mathcal M$. This is the field-theory version of the collective-coordinate construction of Section~\ref{sec:baryons}. Table~\ref{tab:CA} summarises the correspondence.

\begin{table}[h]
\centering
\renewcommand{\arraystretch}{1.25}
\begin{tabular}{lll}
\toprule
\textbf{Current algebra} & \textbf{$U(1)$ gauge theory on $\mathcal M$} & \textbf{Berry} \\
\midrule
$l_0 = \pi - A$ & covariant derivative & adiabatic momentum \\
Schwinger term & field strength $F$ & Berry curvature \\
WZW term & Berry--Wilson loop & Berry phase \\
level $k$ & monopole charge & $2s$ for a spin \\
\bottomrule
\end{tabular}
\caption{Current algebra as a $U(1)$ gauge theory on configuration space.}
\label{tab:CA}
\end{table}

\subsection{The connection in Euler angles}
\label{sec:euler}

To exhibit the monopole explicitly, parametrise $U\in SU(2)$ by Euler angles,
\begin{equation}
U = e^{i\alpha t_3}e^{i\beta t_2}e^{i\gamma t_3},\qquad t_a=\sigma_a/2,\qquad \beta\in[0,\pi],\ \alpha\in[0,2\pi),\ \gamma\in[0,4\pi).
\end{equation}
The WZ three-form is proportional to the Haar measure and has a Dirac-type two-form potential:
\begin{equation}
\frac{ik}{8\pi}\,\sin\beta\;d\alpha\wedge d\beta\wedge d\gamma = dB,\qquad
B = \frac{ik}{8\pi}\,(1-\cos\beta)\;d\gamma\wedge d\alpha .
\end{equation}
The connection on $\mathcal M$ is $A_a(x)=B\big(X_a,\partial_xU\big)$, where $X_aU=it_aU$ are the generating vector fields. Explicitly, with primes denoting $\partial_x$,
\begin{align}
A_1(x)&=-\frac{ik}{8\pi}\,\tan\frac{\beta}{2}\,\cos\alpha\,\big(\alpha'+\cos\beta\,\gamma'\big),\nonumber\\
A_2(x)&=+\frac{ik}{8\pi}\,\tan\frac{\beta}{2}\,\sin\alpha\,\big(\alpha'+\cos\beta\,\gamma'\big),\label{eq:monopole-potential}\\
A_3(x)&=-\frac{ik}{8\pi}\,(1-\cos\beta)\,\gamma' .\nonumber
\end{align}
The potential, evaluated with the assistance of Claude, is regular at $\beta=0$ and singular at $\beta=\pi$, a Dirac string on the circle $U_{11}=0$. The Berry--Wilson-loop phase becomes a local two-dimensional integral over the strip swept out in $(t,x)$:
\begin{equation}
\Phi = \frac{k}{8\pi}\int dt\int_0^{2\pi}dx\;(1-\cos\beta)\,\big(\dot\gamma\,\alpha'-\dot\alpha\,\gamma'\big).
\end{equation}
The potential $B\propto(1-\cos\beta)\,d\gamma\wedge d\alpha$ is the analogue of the Dirac potential $A_\phi\propto(1-\cos\theta)$, with its string on the negative axis. The choice $(-1-\cos\beta)$ moves the string to $\beta=0$; the two choices differ by a gauge transformation on the overlap, so $A$ is a connection on a non-trivial $U(1)$ bundle over $\mathcal M$. The total flux is $2\pi k$, and single-valuedness of $e^{i\Phi}$ requires $k\in\mathbb Z$: Dirac quantisation.

\section{Summary}
\label{sec:outlook}

We summarise the main points.
\begin{enumerate}
\item The ABJ anomaly solved the $\pi^0$ puzzle and counted colours.
\item Anomalies are topological and one-loop exact (Adler--Bardeen), constrained by consistency (Wess--Zumino), equal to the index density of the Dirac operator (Fujikawa; Atiyah--Singer), invariant under the renormalisation group ('t Hooft). Witten's global formulation of the WZ term for the chiral group $SU(N_f)$ led to a quantisation of the coefficient of the WZW term and made an important analogy with a charged particle moving in the field of a magnetic monopole.
\item Integrating out gapped quarks gives the WZW term with universal coefficient $N_c$: it is the phase of $\det D$, cleanly separated from the regulator-dependent modulus~\cite{DW,DSW}.
\item On the collective coordinates of the soliton, the WZW term imposes $Y_R=N_c/3$; the baryon wave functions are monopole harmonics, which fixes their spins, statistics and flavour multiplets~\cite{JW}.
\item The phase of the fermion determinant is a Berry phase. The Schwinger term is the curvature of a $U(1)$ connection on configuration space, and the WZW term is its Berry--Wilson loop~\cite{Wadia2021}.
\item The quantisation of anomaly coefficients is Dirac quantisation of monopoles in field space.
\end{enumerate}

\section{Topics not covered}
\label{sec:notcovered}

The subject of anomalies is vast, and many important developments could not be covered in this article. We list some of them, grouped by theme, with a few words on each.

\paragraph{Mathematics, developed in parallel with the physics.}
\begin{itemize}
\item \emph{The Atiyah--Singer index theorem}~\cite{AS}, announced in 1963 and proved in 1968, at almost the same time as the work of Adler, Bell and Jackiw. We have used only its simplest consequence, that the anomaly is the local index density of the Dirac operator.
\item \emph{The Atiyah--Patodi--Singer theorem}~\cite{APS}, which extends the index theorem to manifolds with boundary and introduces the $\eta$-invariant, a measure of the spectral asymmetry of the Dirac operator. It underlies the modern treatment of global anomalies and of anomaly inflow.
\item \emph{The Witten index}~\cite{WittenIndex}. In a supersymmetric theory with supercharge $Q$ and fermion number $(-1)^F$, the Witten index $\Delta=\Tr\,(-1)^Fe^{-\beta H}$ receives contributions only from the zero-energy states, since non-zero-energy states are paired by $Q$ into bosons and fermions and cancel. It is therefore independent of $\beta$ and invariant under smooth deformations of the theory, exactly like an index, and when it is non-zero supersymmetry is unbroken. For supersymmetric quantum mechanics on a manifold it reduces to the index of the Dirac operator, and to the Euler characteristic in the de Rham case, which gives a supersymmetric proof of the Atiyah--Singer theorem~\cite{AlvarezGaumeSUSY}. It is the quantum-mechanical counterpart of the pairing of non-zero modes by $\gamma_5$ that we used in Section~\ref{sec:uses} to relate the anomaly to the index.
\item \emph{The families index theorem and determinant line bundles}~\cite{AS84,Quillen,BismutFreed}. For a family of Dirac operators parametrised by background fields, the determinant is a section of a line bundle over the parameter space, with a natural connection and curvature. This makes rigorous the Berry connection of the fermion vacuum that we have used informally.
\end{itemize}

\paragraph{The Hamiltonian picture of gauge anomalies.}
In the Hamiltonian formulation, the vacuum of chiral fermions in a background gauge field defines a line bundle over the space of gauge fields modulo gauge transformations. The Gauss-law generators then fail to close, by Schwinger terms~\cite{Faddeev}, and a gauge anomaly appears as a non-trivial Berry curvature or holonomy of this bundle along gauge orbits~\cite{NelsonAG,Sonoda,NiemiSemenoff}. This is the gauge-field counterpart of the construction of Section~\ref{sec:wilson}.

\paragraph{Anomalies across dimensions.}
\begin{itemize}
\item \emph{Descent equations}: the anomaly in $d=2n$ dimensions descends from the characteristic class $\Tr F^{n+1}$ in $d+2$ dimensions through the Chern--Simons form in $d+1$ dimensions. The Wess--Zumino consistency conditions are then automatic, and the WZW term is the Chern--Simons form evaluated on a pure gauge.
\item \emph{Anomaly inflow}~\cite{CallanHarvey}: an anomaly of chiral fermions localised on a $d$-dimensional defect or boundary is cancelled by a $(d+1)$-dimensional Chern--Simons term in the bulk. Witten's five-dimensional disc is then no longer merely a device.
\end{itemize}

\paragraph{Gravity and strings.}
\begin{itemize}
\item \emph{Gravitational anomalies.} Alvarez-Gaum\'e and Witten~\cite{AGW} showed that chiral fields coupled to gravity, spin-$\tfrac12$ and spin-$\tfrac32$ fermions and self-dual tensor fields, can violate general covariance in $d=4k+2$ dimensions, and computed these purely gravitational anomalies. In four dimensions there is no purely gravitational perturbative anomaly, but the mixed gravitational--gauge anomaly, with two graviton vertices and one gauge current, must cancel; in the Standard Model this is the condition $\sum Y=0$ of Table~\ref{tab:SM}. We have not discussed gravitational anomalies further.
\item \emph{The Green--Schwarz mechanism}~\cite{GreenSchwarz}. In ten-dimensional type~I superstring theory, gauge and gravitational anomalies cancel only for the gauge group $SO(32)$, through a counterterm involving the antisymmetric tensor field; the same mechanism works for $E_8\times E_8$. This discovery launched the first superstring revolution.
\item \emph{Global anomalies}, of which Witten's $SU(2)$ anomaly~\cite{WittenSU2} is the prototype: a theory with an odd number of fermion doublets is inconsistent, even though it has no perturbative anomaly.
\end{itemize}

\paragraph{Condensed matter physics and beyond.}
\begin{itemize}
\item \emph{The quantum Hall effect.} Thouless, Kohmoto, Nightingale and den Nijs~\cite{TKNN} showed that the Hall conductance of a filled band is $(e^2/h)$ times the first Chern number of the Berry curvature of the Bloch states over the Brillouin zone. The chiral edge modes carry the $1+1$ dimensional anomaly, cancelled by inflow from the bulk.
\item \emph{Weyl semimetals}, whose Weyl nodes are monopoles of Berry curvature in momentum space, and in which the chiral anomaly leads to observable effects such as negative magnetoresistance; the \emph{chiral magnetic effect}; and \emph{anomalous hydrodynamics}, in which anomalies fix transport coefficients.
\item \emph{The modern view} of 't Hooft anomalies as invertible field theories in one higher dimension, and their role in the study of generalised symmetries.
\end{itemize}

\medskip
It is a pleasure to dedicate this article to Michael Berry, whose geometric phase gave a common language to so many of these ideas.

\section*{Acknowledgements}
I thank Avinash Dhar, R. Shankar and Sanjay Jain for the collaborations on which Sections~\ref{sec:QCD} and~\ref{sec:baryons} are based. This manuscript was carefully read by Avinash Dhar and I thank him for that. I thank Jon Keating, Sthitadhi Roy, Joseph Samuel, Pragya Shukla and Supurna Sinha, the organisers of ``Phases, Caustics and Chaos: Classical and Quantum Aspects'' at ICTS-TIFR, for the invitation to present this talk, and Joseph Samuel for the encouragement to undertake the presentation of the vast topic covered in the talk. I acknowledge the support of the Infosys Foundation Homi Bhabha Chair at ICTS-TIFR.
Claude was used to help evaluate the potential \eqref{eq:monopole-potential}, and to assemble the Latex file.


\end{document}